\documentclass{pasj01}
\usepackage{natbib}
\Received{2019 March 25}
\Accepted{$\langle$acception date$\rangle$}
\Published{$\langle$publication date$\rangle$}

\begin{document}
\title{{}Ly$\alpha$ view around a $z=2.84$ hyperluminous QSO at a node of the cosmic web\thanks{Based on data collected at Subaru Telescope, which is operated by the National Astronomical Observatory of Japan.}}
\author{
Satoshi \textsc{Kikuta},\altaffilmark{1,2,}$^{*}$ 
Yuichi \textsc{Matsuda},\altaffilmark{2,1} 
Renyue \textsc{Cen},\altaffilmark{3} 
Charles C. \textsc{Steidel},\altaffilmark{4} 
Masafumi \textsc{Yagi},\altaffilmark{2,5} 
Tomoki \textsc{Hayashino},\altaffilmark{6}
Masatoshi \textsc{Imanishi},\altaffilmark{2,1} 
Yutaka \textsc{Komiyama},\altaffilmark{2,1} 
Rieko \textsc{Momose},\altaffilmark{7} and
Tomoki \textsc{Saito}\altaffilmark{8}
}
\altaffiltext{1}{Department of Astronomical Science, SOKENDAI (The Graduate University for Advanced Studies), 2-21-1 Osawa, Mitaka, Tokyo 181-8588, Japan}
\altaffiltext{2}{National Astronomical Observatory of Japan (NAOJ), 2-21-1 Osawa, Mitaka, Tokyo 181-8588, Japan}
\altaffiltext{3}{Department of Astrophysical Sciences, Princeton University, Princeton, NJ 08544, USA}
\altaffiltext{4}{Cahill Center for Astronomy and Astrophysics, California Institute of Technology, MS 249-17, Pasadena, CA 91125, USA}
\altaffiltext{5}{Graduate School of Science and Engineering, Hosei University, 3-7-2 Kajinocho, Koganei, Tokyo, 184-8584, Japan}
\altaffiltext{6}{Research Center for Neutrino Science, Tohoku University, Aramaki, Aoba-ku, Sendai 980-8578, Japan}
\altaffiltext{7}{Department of Astronomy, School of Science, The University of Tokyo,
7-3-1 Hongo, Bunkyo, Tokyo 113-0033, Japan}
\altaffiltext{8}{Nishi-Harima Astronomical Observatory, Centre for Astronomy, University of Hyogo, 407-2 Nishigaichi, Sayo-cho, Sayo, Hyogo 679-5313, Japan}
\email{satoshi.kikuta@nao.ac.jp}

\maketitle

\begin{abstract}
We report on the results of deep and wide-field (1.1 deg$^2$) narrow-band observations with Subaru/Hyper Suprime-Cam (HSC) of
a field around a hyperluminous QSO (HLQSO), HS1549+1919, residing in a protocluster at $z=2.84$, to map the large-scale structure of Ly$\alpha$ emitters (LAEs). One HSC pointing enables us to detect 3490 LAEs and 76 extended Ly$\alpha$ blobs (LABs), probing diverse environments from voids to protoclusters. The HLQSO is found to be near the center of the protocluster, which corresponds to the intersection of $\sim$100 cMpc-scale structures of LAEs. LABs are basically distributed along the large-scale structure, with larger ones particularly clustered around the HLQSO, confirming a previously noted tendency of LABs to prefer denser environments. Moreover, the shapes of LABs near the HLQSO appear to be aligned with the large-scale structure. Finally, a deep Ly$\alpha$ image reveals a diffuse Ly$\alpha$ nebula along a filamentary structure with no luminous UV/sub-mm counterpart. 
We suggest that the diffuse nebula is due to a cold filament with high clumping factor illuminated by the QSO, with a required high clumpiness provided by unresolved residing halos of mass $\leq 10^{9-10}M_\odot$.
\end{abstract}

\section{Introduction
\label{sec:intro}}
In the current framework of galaxy formation, 
accretion of cold gas from the intergalactic medium (IGM) through thin filaments (cold streams) is a crucial component which governs subsequent galaxy evolution \citep{Keres2005,Dekel2006,Dekel2009a,Cen2014}.
Accreted gas from the IGM interacts with outflowing gas from galaxies in the circumgalactic medium (CGM, roughly referring to gas within a few times the virial radii of galaxies), regulating the galactic gas supply. Extended Ly$\alpha$ emission, or Ly$\alpha$ blobs \citep[LABs;][]{Steidel2000}, are effective probes for studying the physical properties of the IGM/CGM, particularly for the overdense environment in which they tend to be found \citep{Matsuda2004, Cen2013, Badescu2017}. 
While most LABs have obvious sources for their Ly$\alpha$ emission such as AGNs and starbursts \citep[e.g.,][]{Overzier2013}, some with filamentary morphology might also be related to cold streams \citep{Goerdt2010,Faucher-Giguere2010}. 

\citet{Matsuda2011} suggested a possible relation between LAB morphology and Mpc-scale environments. \citet{Erb2011} reported that six LABs in the HS1700+643 protocluster at $z=2.3$ \citep{Steidel2005} are distributed along two linear structures with position angles of each oriented along the same lines. Although these results suggest a link between LABs and the large-scale structure (LSS), 
whether or not this is a general trend is still not known due to a lack of very deep and wide observations.

In this Letter, we present a new example of such a field suggesting a correlation between the LSS and Ly$\alpha$ emission, around a hyperluminous QSO (HLQSO) HS1549+1919 at $z=2.84$. Previous galaxy redshift surveys have revealed a massive protocluster around the HLQSO \citep{Steidel2011,Trainor2012,Mostardi2013}. With new wide-field imaging data obtained with the Hyper Suprime-Cam \citep[HSC; $\phi=1.5\deg$,][]{Miyazaki2012a} on the 8.2m Subaru Telescope, we were able to trace structures on $>100$ comoving Mpc (cMpc) scales around the HLQSO. Additionally, the HSC images led us to discover Mpc-scale Ly$\alpha$ emission surrounding the HLQSO, which may be enhanced due to the unusual QSO activity within a small region.
Throughout the paper we use the AB magnitude system and a cosmology with $\Omega_{\mathrm{m}}=0.3$, $\Omega_{\mathrm{\Lambda}}=0.7$, and $H_0=70 \mathrm{~km ~s^{-1}~Mpc^{-1}}$. At $z=2.84$, 1\arcsec and 1\arcmin corresponds to 7.8 and 470 physical kpc (pkpc), respectively.

\section{Data and Analyses
\label{sec:obs}}
We observed the field centered on HS1549+1919 (15{\mbox{$~\!\!^{\mathrm h}$}}51{\mbox{$~\!\!^{\mathrm m}$}}52{\mbox{$.\!\!^{\mathrm s}$}}47, 
+19\arcdeg11\arcmin04{\mbox{$.\!\!\arcsec$}}0 (J2000.0), $z=2.843$, $L_\mathrm{1450}=1.5\times10^{14}\LO$, \citealt{Trainor2012}), with Subaru/HSC on 2016 June 2--3 (UT; Program ID: S16A-110, PI: Y. Matsuda). 
We used a g-band ($\lambda_\mathrm{c}=4712$\AA, $\Delta \lambda=1479$\AA) and a narrow-band filter NB468 ($\lambda_\mathrm{c}=4683$\AA, $\Delta \lambda=88$\AA). We took images with short exposure time (20 sec for g and 200 sec for NB468) to avoid saturation of the HLQSO and with 5-point dithering (RDITH$=$600\arcsec) and 12 different position angles. We stacked, respectively, 389 and 113 individual exposures (total of 2.2 hrs and 6.3 hrs, respectively) to produce final g-band and NB468 images. 

The raw data were reduced with HSC pipeline\footnote{https://hsc.mtk.nao.ac.jp/pipedoc\_e/} version 4.0.5 \citep{Bosch2018}, using additional packages for ghost and satellite masking\footnote{https://hsc.mtk.nao.ac.jp/pipedoc\_e/e\_tips/ghost.html} and global sky subtraction\footnote{https://hsc.mtk.nao.ac.jp/pipedoc\_e/e\_tips/skysub.html\#global-sky}, which estimates and subtracts the sky on scales larger than that of individual CCDs in the mosaic with a grid size of 6000 pixel (17\arcmin). 
Astrometry and photometric zero points were calibrated using Pan-STARRS1 catalogs \citep{Schlafly2012}. 
The FWHMs of stellar sources are 0{\mbox{$.\!\!\arcsec$}}77 and 0{\mbox{$.\!\!\arcsec$}}65 in g-band and NB468, respectively. The NB468 image was smoothed to match the FWHM of stellar sources in the g-band image. After correction for Galactic extinction of $A_\textrm{g}=A_\textrm{NB468}=0.17$ mag \citep{Schlegel1998}, a 5$\sigma$ limiting magnitude measured with 1{\mbox{$.\!\!\arcsec$}}5 diameter aperture is 27.4 (26.6) mag for g-band (NB468) image. As the S/N ratio near the edge of the images becomes lower, we only use a $72\arcmin$ diameter circular region centered at the HLQSO for the analysis below. 

We used photometric criteria below to select LAE at $z=2.815$--2.887:
\begin{eqnarray}
\mathrm{g} - \mathrm{NB468} & > & \max(0.5, 0.1+4\sigma(\mathrm{g}-\mathrm{NB468})) \label{eq:1}\\
\mathrm{NB468} & < & 26.6 \label{eq:2}
\end{eqnarray}
where $\sigma(\mathrm{g}-\mathrm{NB468})$ denotes the expected 1$\sigma$ deviation of the quantity 
$\mathrm{g}-\mathrm{NB468}$ for a flat continuum ($f_\nu=$const.) source. 
The criterion $\mathrm{g} - \mathrm{NB468} > 0.5$ corresponds to $\mathrm{EW_{obs}}> 59$\mbox{\AA}. 
For objects fainter than $2\sigma$ in the g-band, we replace their g-band magnitude by
their $2\sigma$ limiting magnitude as a lower limit. SExtracor \citep{Bertin1996} was used to perform 1{\mbox{$.\!\!\arcsec$}}5 aperture photometry with double-image mode, with the NB468 image used as the detection band. We set a background mesh size of 64 pixels ($=11$\arcsec) to estimate local sky values. Spurious detections such as diffraction spikes around bright stars and cosmic rays are excluded. As a result, we detected 3490 LAEs within $r<36\arcmin$ from the QSO position. 

We also selected extended LAEs or Ly$\alpha$ blobs (LABs) using an additional criterion on their isophotal areas on the Ly$\alpha$ image. A pure Ly$\alpha$ image was created by subtracting the g-band image from the NB468 image after scaling by their relative zero points. The g-band overlaps with the NB468 and this has the slight effect ($\sim6\%$ at the NB center) of an oversubtraction. We smoothed the Ly$\alpha$ image to reduce noise with gaussian kernel $\sigma=3$ pixel or 0{\mbox{$.\!\!\arcsec$}}5 (FWHM 1{\mbox{$.\!\!\arcsec$}}2). LABs were selected as objects which satisfy Equation (\ref{eq:1}) and (\ref{eq:2}), and in addition have 2$\sigma$ (28.36 mag arcsec$^{-2}$) isophotal areas larger than 16 arcsec$^2$ in the Ly$\alpha$ image. For those extended sources, we used the isophotal area to calculate the color, and a background mesh size of 176 pixels ($30\arcsec$) was used for local sky estimation. These criteria were chosen to be consistent with those used by \citet{Matsuda2004}. After rejecting obviously spurious detections, we detected 76 LABs in the field.

\section{Results}
\subsection{Distribution of LAEs and LABs
\label{sec:laelab}}
\begin{figure}
  \begin{center}
   \includegraphics[width=0.48\textwidth]{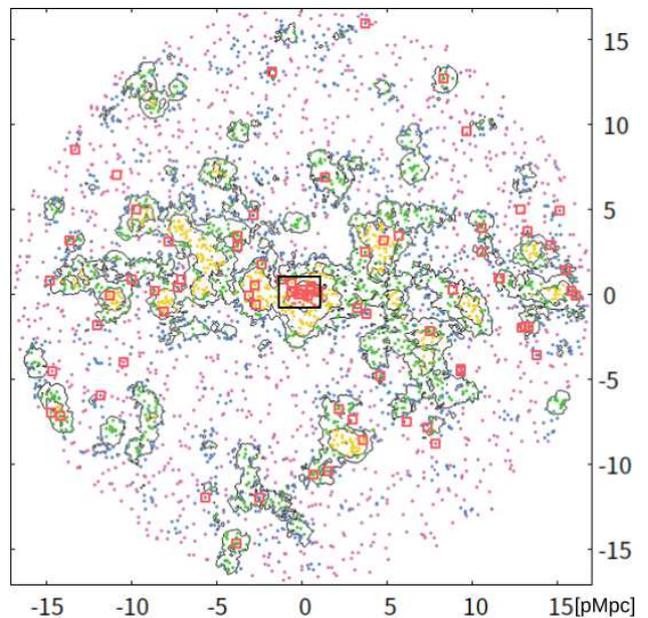}
  \end{center}
  \caption{Spatial distribution of LAEs at $z=2.84$ with north up and east left. The origin (0,0) is set to the QSO location. Colored points denote selected LAEs, with their colors reflect local environments they reside (see Figure \ref{fig:env}). Red squares denote LABs. Black contours are LAE overdensity $\delta=0.3, 1, 2.5$. Black rectangle shows the area highlighted in Figure \ref{fig:halo}. 5pMpc corresponds to an angular scale of 10{\mbox{$.\!\!\arcmin$}}7.
\label{fig:maps}}
\end{figure}

Figure \ref{fig:maps} shows the spatial distribution of LAEs/LABs around the HLQSO, overlaid on the number density map of LAEs measured with the ``fixed aperture method'': 
we measured a local density of LAEs by counting their number $n$ within a fixed aperture. Then we calculated the average $\bar n$ and overdensity $\delta$ as $\delta\equiv(n-\bar n)/\bar n$. The value of $\delta$ depends strongly on the (arbitrarily chosen) aperture radius. Here we used an aperture radius of 1{\mbox{$.\!\!\arcmin$}}8 (or 0.83 pMpc) to be consistent with the measurement of \citet{Matsuda2012} and \citet{Yamada2012}. 

\begin{figure}
  \begin{center}
   \includegraphics[width=0.45\textwidth]{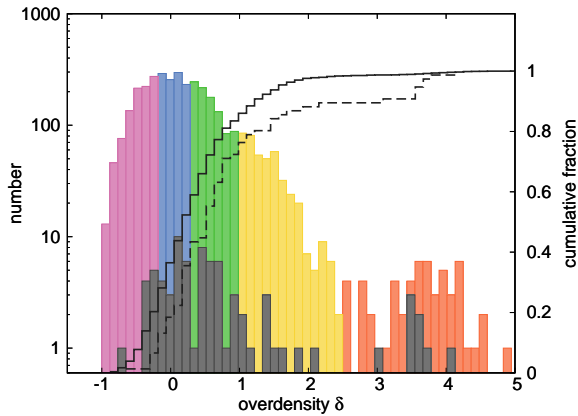}
  \end{center}
  \caption{Distribution of $\delta$. Color (gray) boxes indicate the distribution for LAEs (LABs). Solid line shows a cumulative distribution function of $\delta$ for LAEs while dashed line shows that for LABs, which is clearly skewed toward the right-hand (thus denser) side.
\label{fig:env}}
\end{figure}

The HLQSO is located at the origin in Figure \ref{fig:maps}. Two arm-like structures extend toward the north-east and north-west, revealing that the HLQSO lies near their intersection. 
LABs are distributed along the structure, avoiding the huge void to the south of the protocluster. 
The distribution of $\delta$ for LAEs and LABs is presented in Figure \ref{fig:env}. 
As previously noted, a clear trend of LABs favoring denser environments is seen in both Figure \ref{fig:maps} and in the cumulative distribution function in Figure \ref{fig:env}. This trend becomes even clearer for larger LABs; of eight LABs with isophotal area $>50$ arcsec$^2$, four are located within 1.5 pMpc (see Figure \ref{fig:halo}) of the HLQSO, and one lies only 2.7 pMpc away. 
The number density of LABs averaged over the entire field is $7.8\times10^{-5}\mathrm{~cMpc^{-3}}$ using the survey volume (dictated by the field of view and width of the NB468, $\Delta z=0.075$ or equivalently $d_\mathrm{C}=73$ cMpc) of $\sim10^6 \mathrm{~cMpc^{3}}$ without any correction. This is an order of magnitude higher than that observed in random fields at $z=2.3$ \citep{Yang2010}, largely because of the difference in sensitivity. 
A survey by \citet{Matsuda2004} targeting the SSA22 protocluster at $z=3.1$ found 35 LABs within a $31\arcmin\times23\arcmin$ field ($1.3\times10^5\mathrm{~cMpc^{3}}$) with similar sensitivity and criteria to ours, resulting in $\sim2.7\times10^{-4}\mathrm{~cMpc^{-3}}$. If we use the same $31\arcmin\times23\arcmin$ window centered at the QSO location, we obtain $\sim1.4\times10^{-4}\mathrm{~cMpc^{-3}}$ for the HS1549 field. Note, however, these values are sensitive to the window size and the extent of the protoclusters. Within a 4 pMpc diameter aperture \citep[a typical size for protoclusters at $z\sim3$, ][]{Chiang2013}, the number density is a factor of 5 higher.

\subsection{Ly$\alpha$ emission around the HLQSO \label{sec:lyaimage}}
\begin{figure*}
  \begin{center}
   \includegraphics[width=0.75\textwidth]{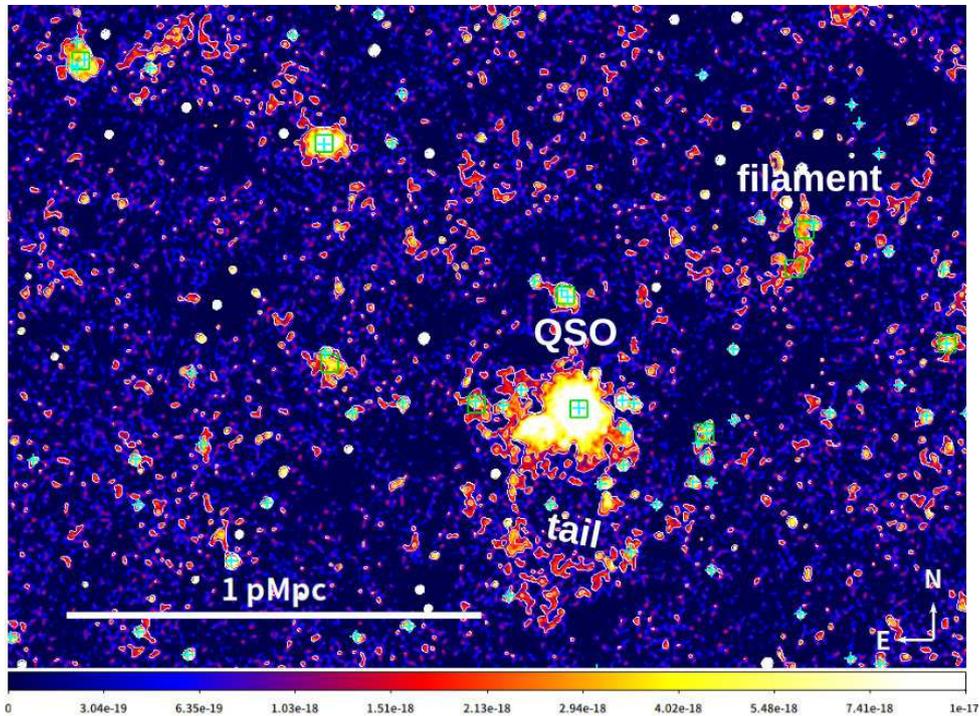}
  \end{center}
  \caption{Cutout Ly$\alpha$ image showing the 5\arcmin $\times$3{\mbox{$.\!\!\arcmin$}}5 area around the HLQSO with north up and east left. Cyan crosses (green squares) denote objects selected as LAEs (LABs). Color bar indicates Ly$\alpha$ brightness in $\mathrm{erg~s^{-1}~cm^{-2}~arcsec^{-2}}$ using an asinh stretch. White contours correspond to $1\times10^{-18}$ and $2\times10^{-18}~\mathrm{erg~s^{-1}~cm^{-2}~arcsec^{-2}}$ (respectively correspond to $\sim1\sigma$ and $2\sigma$). 
\label{fig:halo}}
\end{figure*}

A close-up smoothed Ly$\alpha$ image around the HLQSO is shown in Figure \ref{fig:halo}, using a background subtraction mesh size of 30\arcsec (235 pkpc)\footnote{We confirmed nebulae discussed below persist with larger mesh sizes of 1\arcmin and 2\arcmin. This is also checked with two images made by splitting 113 NB468 exposures into two and separately stacking them. Also, we checked no dubious negative pattern exists at their location in the g-band image which could produce spurious nebulae in the Ly$\alpha$ image.}. The HLQSO is the brightest source near the center. An enormous Ly$\alpha$ nebula around the QSO reported in \citet{ChristopherMartin2014a} is clearly seen. 
Although not connected to this main nebula, there is a chain of diffuse Ly$\alpha$ emission toward the south (marked as ``tail''), comprising a Mpc-scale Ly$\alpha$ structure. No source is detected as an LAE within the tail except for the one at the tip. We caution that, however, some parts of the tail are only significant at 1-2$\sigma$ levels. 

In this $5\arcmin\times3{\mbox{$.\!\!\arcmin$}}5$ image, 10 sources are detected as LABs (green squares), most of which have a size of $>100$ pkpc. The distribution and shapes of LABs appear to broadly follow the LSS toward north-east and north-west in Figure \ref{fig:maps}, although the alignment is not as obvious as that described by \citet{Erb2011}. Similar to the tail, diffuse emission to the north-west of the HLQSO (marked as ``filament'')\footnote{These are originally selected as two separated LABs. We regard it as one hereafter.} has no LAE counterpart except for the one at the north end, despite its total Ly$\alpha$ luminosity of $\sim10^{43.3}\mathrm{~erg~s^{-1}}$. If powered by star formation, $\mathrm{SFR}=12~\MO\mathrm{~yr^{-1}}$ is required, using calibration of \citet{Murphy2011} and $\mathrm{Ly\alpha}/\mathrm{H\alpha}=8.7$. There are two UV sources with $\mathrm{g}=25.6$ and 25.8 mag with $\mathrm{g}-\mathrm{NB468}\sim0$ just outside of the 2$\sigma$ SB contour. Even if they are both at the same redshift as the filament, their combined SFR amounts to only $\sim7~\MO\mathrm{~yr^{-1}}$.

\section{Discussion
\label{sec:discussion}}
To estimate descendant halo mass of this protocluster, 
we compare our results with those of \citet{Chiang2013}, who provided the relation between overdensity $\delta$ of galaxies at $z=3$ and descendant cluster total mass 
(see their Figure 13). We used a circular aperture radius of 8.46 cMpc and LAEs with $L_\mathrm{Ly\alpha}>10^{42}\mathrm{~erg~s^{-1}}$ to match the area of square aperture they provided ($15\times15$ cMpc$^2$) and their SFR threshold of $>1~\MO\mathrm{~yr^{-1}}$.
In our case, we measured peak value of $\delta_\mathrm{8.46cMpc}=2.0$ with the redshift uncertainty of LAEs of $\Delta z=0.075$,  
making the HS1549 field a highly reliable progenitor of Coma-type ($M_\mathrm{halo}>10^{15}\MO$) cluster. A similar conclusion is reached by examining the overdensity of continuum-selected galaxies in the KBSS spectroscopic sample (C. Steidel, et al. 2019, in prep.). At $z\sim3$, the most massive halos could have $M_\mathrm{halo}\sim10^{13}\MO$. These redshift and halo mass are respectively lower and upper limit where cold streams can penetrate through growing hot media \citep{Dekel2006}, possibly enabling extreme starburst and SMBH activity around the QSO. 

It is interesting to consider the origin of the two mysterious nebulae, namely the ``tail'' and ``filament'' in Figure \ref{fig:halo}. 
The overall shape of the tail is reminiscent of tidal features in nearby galaxies, albeit an order of magnitude size difference. They could be remnant features of mergers and interactions with the main halo which are distributed prior to the onset of the HLQSO. 
Another possibility is that they trace cold streams or halos of unresolved (proto)galaxies embedded within the LSS. This scenario seems incompatible for the tail with its large curvature and the fact that there is no structure of LAEs to the south. 
By contrast, the thin shape, orientation pinpointing the HLQSO, and alignment with the LSS of the ``filament'' is consistent with expectations for cold streams along the LSS. It is located at $\sim650$ pkpc (projected, $\sim3\times R_\mathrm{vir}$) away from the QSO. One mystery then arises is the absence of emission between the QSO and the filament. 
This absence may be explained by a variability, anisotropy of QSO illumination and/or a local lower density region,
which then simultaneously provides the required illumination of the filament by the QSO discussed in the following paragraph.
We thus suggest that the filament may represent a cold stream
that has been long sought after. 

The configuration and position angles of other LABs are also roughly aligned with the LSS. Specifically, 5 LABs to the north and east of the QSO showed $\cos\theta>0.8$, where $\theta$ is the angle between their major axis and the line to the QSO, while if their orientations are randomly distributed there is 1\% chance for this to happen. 
This suggests enhanced cold gas accretion and interactions along the structure efficiently distribute gas around galaxies and activate AGNs/starbursts which light up the gas, making it visible as LABs. Indeed, three brightest LABs in Figure \ref{fig:halo} (the HLQSO itself and the two to the north-east) are associated with either SMG \citep[the eastmost one,][]{Lacaille2018} or QSOs, with three more AGNs within just 150 pkpc from the HLQSO, suggesting enhanced interactions (C. Steidel et al. 2019, in prep.). 

Because no counterpart is identified within the filament at present, the origin of the Ly$\alpha$ emission is ambiguous --- possibilities include obscured galaxies, gravitational cooling, scattering of Ly$\alpha$ photon from the HLQSO, or photoionization by ionizing photons from the HLQSO \citep[``QSO fluorescence'', e.g., ][]{Kollmeier2010}. \citet{Lacaille2018} conducted JCMT/SCUBA2 observations with 0.6 mJy beam$^{-1}$ sensitivity at 850 $\mu$m (corresponding to $\mathrm{SFR}\sim45~\MO\mathrm{~yr^{-1}}$) for this field. They found a source with a few mJy at 850 $\mu$m near the filament, but it is at $\sim10\arcsec$ away and at present has no redshift information, leaving it unconvincing as a power source. Fainter SMGs can still explain observed $L_\mathrm{Ly\alpha}$ and need to be constrained in the future. The contribution of gravitational cooling to the Ly$\alpha$ emission, though still difficult to predict due to its exponential dependence on temperature, can in principle be large enough to be detected \citep[e.g.,][]{Goerdt2010,Rosdahl2012} but only within $\sim R_\mathrm{vir}$ from the massive ($M_\mathrm{halo}>10^{12}$) halo center in most cases,
hence this mechanism is unlikely at the location in question. Considering the luminosity of the HLQSO and assuming the gas sees the QSO radiation, hydrogen in the filament should be almost ionized. Thus, the most promising power source at this stage is QSO fluorescence. 

We crudely estimate the plausibility of this scenario. The ionizing photon production rate from the HLQSO is estimated as $\dot{N}=10^{58}~\mathrm{s^{-1}}$ using the observed UV luminosity of $\nu L_\mathrm{\nu,1450\AA}=1.5\times10^{14}\LO$ and assuming UV continuum slope of $\nu^{-1}$. Modeling the filament as a column with a height of 200 kpc and a diameter of 50 kpc, the amount of ionizing photon available to the filament at the distance of 650 kpc is $\Omega/(4\pi)\cdot\dot{N}=2\times10^{56}~\mathrm{s^{-1}}$ 
or equivalently  $L_\mathrm{Ly\alpha}=2\times10^{45}~\mathrm{erg~s^{-1}}$, which can power the observed luminosity of the filament, $L_\mathrm{Ly\alpha}=2\times10^{43}~\mathrm{erg~s^{-1}}$.
The predicted HI fraction can be written as $C n_\mathrm{e} \alpha/\Gamma=4\times 10^{-4}Cn_\mathrm{e}$, where $C$ is the clumping factor, $n_\mathrm{e}$ is the electron density [$\mathrm{cm^{-3}}$], and $\alpha=2.6\times10^{-13}\mathrm{~cm^3~s^{-1}}$ denotes the case-B recombination rate at temperature of $10^4$ K, and $\Gamma$ denotes the photoionization rate [$\mathrm{s^{-1}}$]. Under these assumptions, the luminosity of the filament will be roughly $L_\mathrm{Ly\alpha}=0.68h\nu_\mathrm{Ly\alpha}\times C n_\mathrm{e}^2 \alpha V=3\times10^{46}Cn_\mathrm{e}^2~\mathrm{erg~s^{-1}}$, where $h\nu_\mathrm{Ly\alpha}$ is the energy of Ly$\alpha$ photon and $V$ is the assumed total volume of the structure. As an example, $C\sim10^3$ with a density $n_\mathrm{e}\sim10^{-3}~\mathrm{cm^{-3}}$,
corresponding to about 100 times the mean baryon density at $z=2.84$, could explain the luminosity of the filament.

How is this high clumping factor achieved? We advocate a mechanism where the high clumpiness of gas in the filament is due to the CGM of a significant number of faint halos of mass in the range of $\leq10^{9-10}\MO$. Regions just around self-shielded spheres where density is sufficiently high are ionized by the QSO and emit Ly$\alpha$ photons which are then scattered by residual HI gas in the filament. We will detail this model in a separate paper. 
Additionally, the sub-kpc scale structure in the CGM/IGM \citep[e.g.,][]{Rauch1999,Schaye2007,Hennawi2015,Cantalupo2019} may drastically change its emission properties \citep[e.g.,][]{Corlies2018}. 
Current cosmological simulations lack sufficient resolution to resolve such scales. 
These small-scale clumps, if proven to exist in substantial abundance even in the IGM, may change the above scenario.

\section{Summary\label{sec:summary}}
We have reported the results of our wide-field Ly$\alpha$ imaging around the HLQSO HS1549+1919 at $z=2.84$ with Subaru/HSC. 
We were able to map the large-scale structure traced by Ly$\alpha$ emission on $>100$ cMpc scales and probe diverse environments from voids to protoclusters within a single pointing of HSC, clearly displaying its power (Figure \ref{fig:maps}). Furthermore, we identified a variety of Ly$\alpha$ nebulae near the central part of the protocluster centered on the HLQSO. What especially stand out in this field are the overdensity of LABs, their apparent alignment with that of the large-scale structure, and the existence of a ``filament'', that may be ambient gas illuminated by the HLQSO (Figure \ref{fig:halo}). 
These results suggest enhanced interactions which redistribute gas around galaxies and abundant cold gas in the densest environments at cosmic noon.  
With a help of QSO fluorescence, more LABs similar to the filament will be found. A detailed comparison with integral field spectroscopy datacubes of such LABs and numerical simulations will clarify their true nature and shed light on crucial gaseous phenomena which drive galaxy evolution.

\begin{ack}
We thank Yukie Oishi and the HSC pipeline team for their helpful comments on HSC data analyses, the anonymous referee for his/her thoughtful comments, and Pimpunyawat Tummuangpak for reading the manuscript. 
We would like to acknowledge all who supported our observations with the Subaru Telescope, including the staffs of the National Astronomical Observatory of Japan, Maunakea Observatories, and the local Hawaiian people who have been making efforts to preserve and share the beautiful dark sky of Maunakea with us. 

Data analysis was carried out on the Multi-wavelength Data Analysis System operated by the Astronomy Data Center (ADC), National Astronomical Observatory of Japan.

SK acknowledges supports from the JSPS KAKENHI grant No. 18J11477 and the Course-by-Course Education Program of SOKENDAI.
YM acknowledges supports from the JSPS KAKENHI grant No. 25287043, 17H04831, 17KK0098.
MI acknowledges supports from the JSPS KAKENHI grant No. 15K05030.
\end{ack}

\end{document}